\documentclass[aps, twocolumn,noshowpacs,preprintnumbers,amsmath,amssymb]{revtex4}

\usepackage{verbatim}
\usepackage{wasysym}

\usepackage{graphicx}
\usepackage{dcolumn}
\usepackage{bm}
\usepackage{wrapfig}

\begin{document}

\title{Tunable electrical transport through annealed monolayers of monodisperse cobalt-platinum nanoparticles}

\author{Yuxue Cai}
\author{Denis Wolfkuhler}
\author{Anton Myalitsin}
\author{Jan Perlich}
\author{Andreas Meyer}
\author{Christian Klinke}
\email{klinke@chemie.uni-hamburg.de}
\affiliation{Institute of Physical Chemistry, University of Hamburg, 20146 Hamburg, Germany}

\begin{abstract} 

We synthesized monodisperse cobalt-platinum nanoparticles Co$_{0.14-0.22}$ Pt$_{0.86-0.78}$ of 9 nm in diameter by colloidal chemistry methods and deposited them by the Langmuir-Blodgett technique as highly ordered monolayers onto substrates with e-beam defined gold electrodes. Upon annealing we observe an increase of conductivity over more than 4 orders of magnitude. A first attempt of explanation of this unanticipated effect, a nanoparticle displacement, could not be confirmed for annealing temperatures below 400$^{\circ}$C. A second approach, a carbonization of the ligands, however, could be confirmed by Raman spectroscopy. The simple thermal treatment allows tuning essential properties of electronic devices based on nanoparticles by the manipulation of the interparticle coupling, namely the electrical conductivity, the Coulomb blockade characteristic, and the activation energy of the system. 

\end{abstract}

\maketitle

Metallic nanoparticles are receiving increased attention due to their extensive application in the fields of electronics, catalysis, high-density magnetic recording media, and biotechnology \cite{1,2,3,4}. Monodisperse nanocrystals of magnetic CoPt and FePt alloys are among the candidates for applications in ultra-high-density storage devices due to their large uni-axial anisotropy and high chemical stability \cite{5}. An important step for further commercial exploitation of nanoparticles is their assembly into topologically predefined superstructures and their packing. The collective properties of assemblies of nanoparticles interacting with each other can differ from the behavior of individual particles, with amazing application potential e.g. in optical and electronic devices \cite{6}. The transport properties and mechanisms of 2D metallic nanoparticle arrays are discussed e.g. by the Jaeger group \cite{7,8,9}.

In previous publications we reported about highly ordered, monolayered cobalt-platinum nanoparticle films and their electrical properties \cite{10,11}. Such films could find applications e.g. as sensors \cite{12,13}. For such application it is desirable to be able to tailor the electrical properties such as conductivity, resulting in a tunable signal-to-noise ratio. The nanoparticle film conductivity can be improved decisively by doping semiconductor particles \cite{14,15}, by changing the chemical environment around the particles by introducing conductive ligands \cite{16,17,18}, or by ligands removal \cite{19,20,21,22}. For the latter plasma treatment \cite{23} and annealing are applicable \cite{24}. The thermal stability and especially the annealing behavior of CoPt nanoparticles were already investigated \cite{25,26}. Multilayers of particles were heated in air or under vacuum in order to evaluate the particle stability and to find the transition temperature from a disordered to an ordered fcc crystal structure. 

In the following we discuss the effect of annealing on the conductivity of monolayers of CoPt nanoparticles. Upon annealing, we notice a conductivity increase of several orders of magnitude with rising annealing temperature, until a maximum is reached at about 400$^{\circ}$C. Since this result was unexpected for monolayer films, we investigated the occurring changes in more detail and developed two explanations for our observations. The first possibility is a moderate nanoparticle displacement resulting in preferred paths for the electron transport due to reduced tunnel barriers on some paths. As a consequence the conductivity is decreased in other film areas nevertheless giving elevated overall film conductivity due to the exponential relationship between the tunnel resistance and interparticle distance (Fig. 1B). Another explanation is a thermally induced carbonization of ligands, triggered by the catalytically active CoPt alloy (Fig. 1C). The ligand shells around the particles are saturated carbon chains (1-adamantanecarboxylic acid and hexadecylamine). During the annealing procedure a partial oxidation of the hydrocarbon chains might occur on the CoPt surface resulting in conductive sp$^{2}$-hybridized carbon atoms.

\begin{figure*}[htbp]
  \centering
  \includegraphics[width=0.98\textwidth]{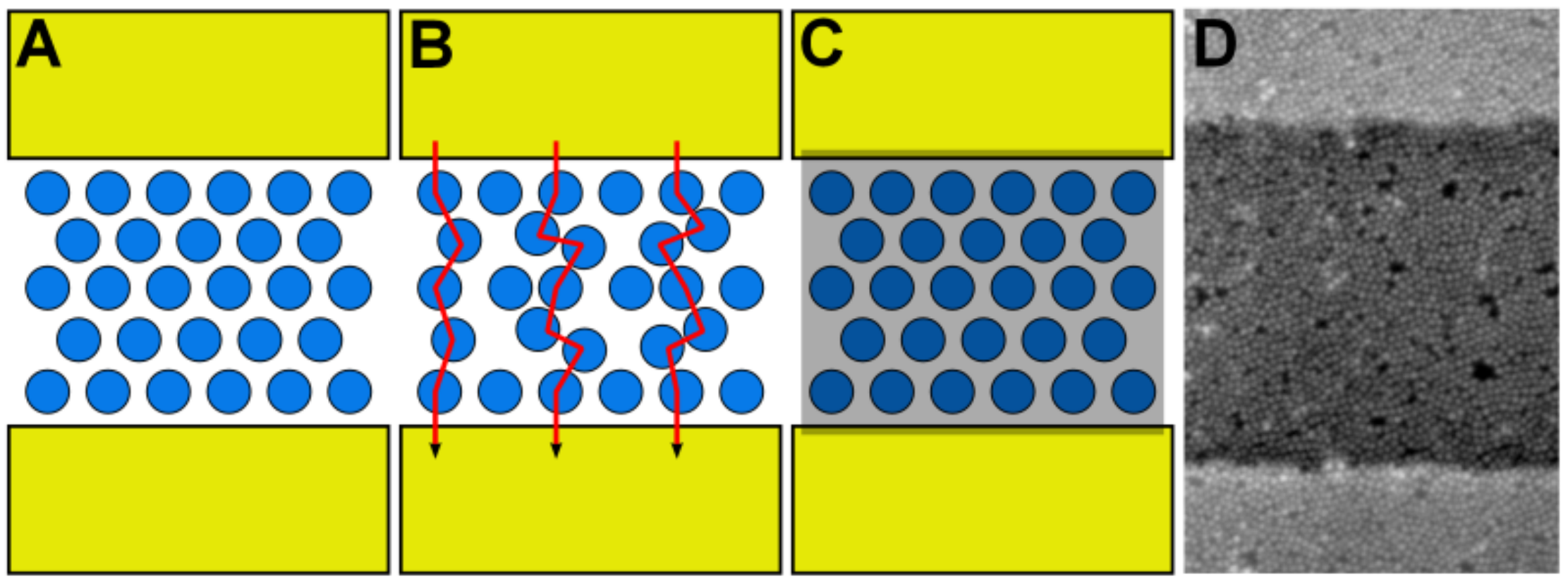}
  \caption{\textit{Possible mechanisms. A) Pristine, well ordered sample, B) first possible conduction mechanism after annealing: shift of nanoparticles and enhanced conductivity through paths with shorter tunnel barriers, C) second mechanism: carbonization of ligands. D) SEM image of a CoPt nanoparticle monolayer deposited on a substrate structured with gold electrodes (the brighter parts).}}
\end{figure*}

\subsection*{RESULTS AND DISCUSSION}

The used CoPt nanoparticles were synthesized following a procedure introduced by Shevchenko et al. \cite{26}. The obtained nanoparticles had a diameter of 9.0 +/- 0.6 nm. All nanoparticles had an fcc crystal structure. The composition of Co$_{0.14-0.22}$ Pt$_{0.86-0.78}$ was determined by energy dispersive X-ray analysis (EDX). After a washing procedure, monolayers of CoPt nanoparticles were prepared by the Langmuir-Blodgett method as described in previous publications of our group \cite{10,11}. For electrical transport measurements the monolayers were deposited on silicon dioxide wafers with gold electrodes with a width of 1 µm and an inter-electrode distance of 1.2 $\mu$m. 

In order to probe the transport properties of the nanoparticle films, we measured the dc response of the samples, which were annealed for 30 min under vacuum at increasing temperatures (for annealing kinetics see SI). Fig. 2A shows I-V curves of a device annealed in steps from room temperature to 500$^{\circ}$C. Fig. 2B shows the conductivity at 3 V plotted against the annealing temperature. The conductivity increases with each annealing step. First we observed a steep increase over several orders of magnitude which levels off at higher annealing temperatures. At annealing temperatures beyond 400$^{\circ}$C the conductivity drops significantly, below the conductivity of the untreated film.

\begin{figure}[htbp]
  \centering
  \includegraphics[width=0.45\textwidth]{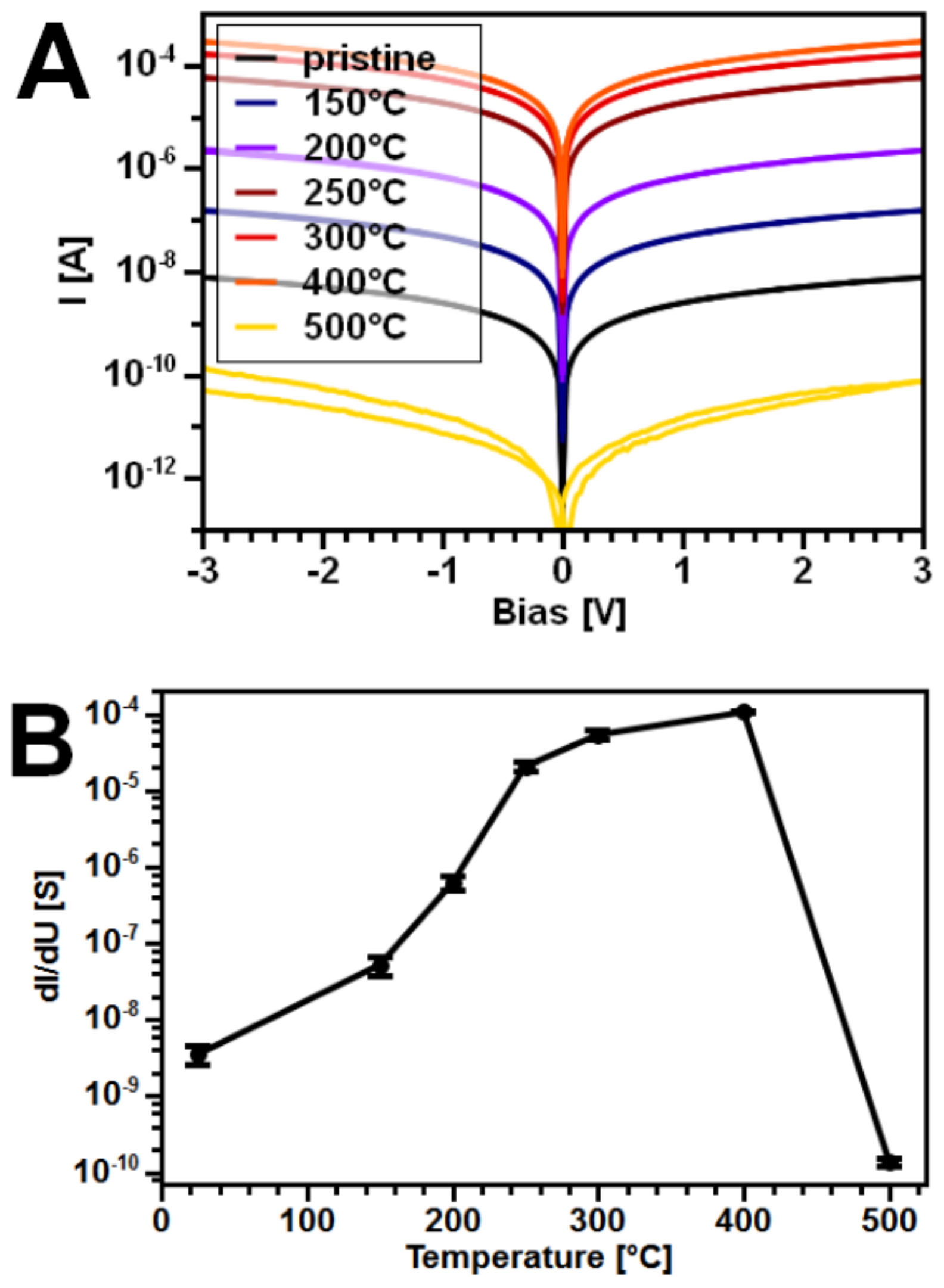}
  \caption{\textit{A) I-V curves of a monolayer film, annealed at different temperatures and measured at room temperature. The conductivity increases with increasing annealing temperature up to 400$^{\circ}$C. B) The conductivity at 3 V plotted against the annealing temperature.}}
\end{figure}

In order to analyze the reason for the pronounced conductivity increase after the thermal treatment up to 400$^{\circ}$C and for the conductivity drop above 400$^{\circ}$C, we examined the samples before and after every annealing step by scanning electron microscopy (SEM), transmission electron microscopy (TEM), and grazing incidence small-angle x-ray scattering (GISAX). With all three structural analysis methods we could not see any substantial change in the samples below 400$^{\circ}$C. The SEM images in Fig. 3A show that the particles are well ordered and packed in a hexagonal arrangement before heating. Annealing at 400$^{\circ}$C resulted in rifts in the monolayer films (Fig. 3B). At 500$^{\circ}$C and higher the particles start to agglomerate (Fig. 3C). This effect is also seen in the TEM images (Fig. 3D-F). At an annealing temperature above 400$^{\circ}$C the particles start to form small agglomerates consisting of a few particles. The electron diffraction shows no crystal lattice difference between thermally treated and untreated particles. The agglomerates are bigger crystals with the same fcc lattice like the untreated particles. These studies show that heating of CoPt nanoparticle films up to 400$^{\circ}$C can be used to increase the conductivity of the film without loss of the film morphology.

\begin{figure*}[htbp]
  \centering
  \includegraphics[width=0.98\textwidth]{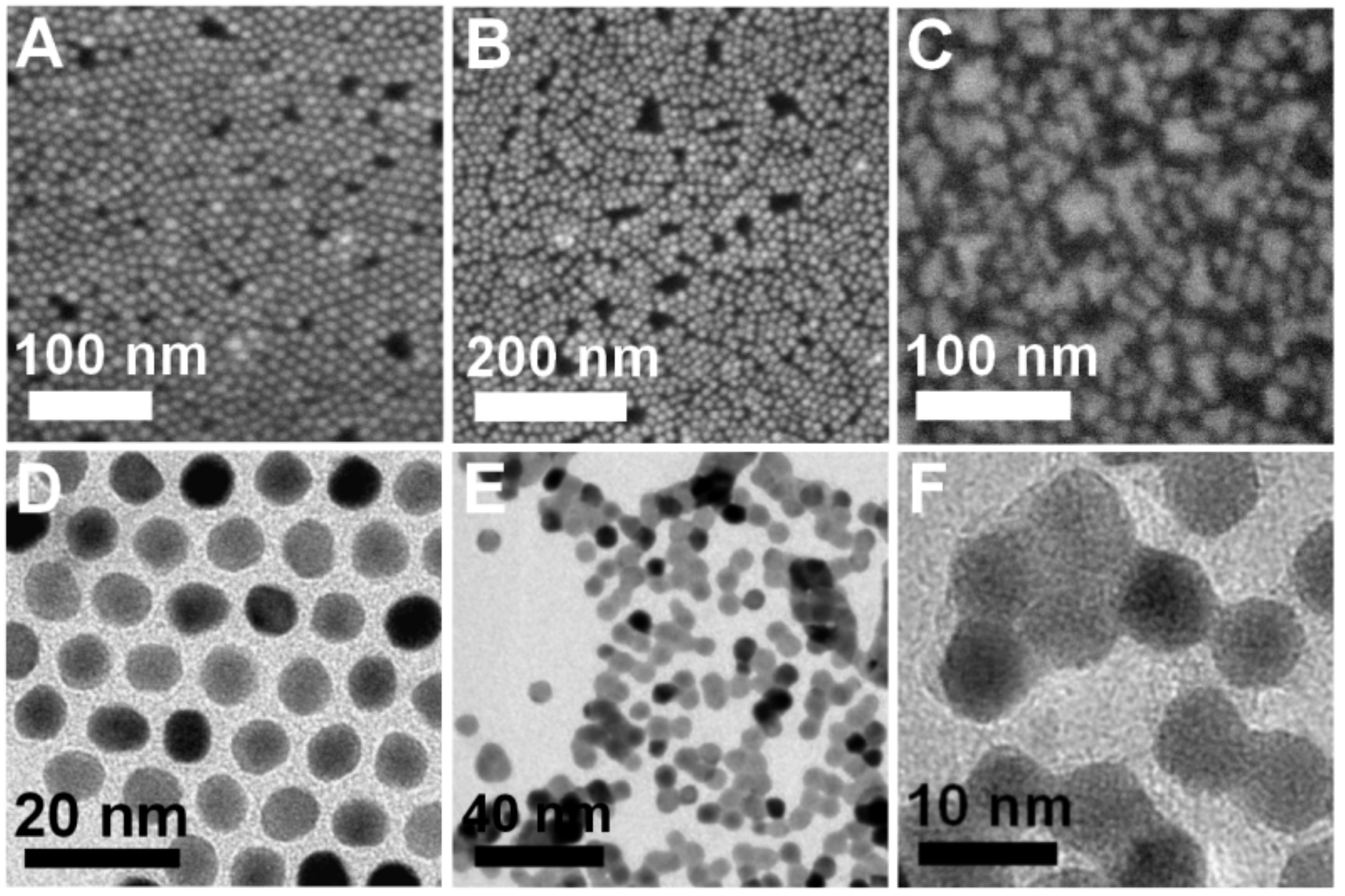}
  \caption{\textit{SEM images of monolayer films A) pristine, B) annealed at 400$^{\circ}$C and, C) at 500$^{\circ}$C. The annealed films show rifts and agglomerates. D) TEM image of as-synthesized nanoparticles, E) TEM images of particles annealed at 400$^{\circ}$C, F) higher magnification.}}
\end{figure*}

While the short-range particle ordering was observed by microscopic techniques we used GISAXS measurements to examine long range properties. From simulations of GISAXS patterns it is possible to get information about the form factor and the interference function, which leads to the size and shape of the particles and their lateral distances \cite{27}. We do not observe any structural changes in the films up to an annealing temperature of 350$^{\circ}$C. We notice a significant change in the scattering pattern after annealing the film to 400$^{\circ}$C or above (Fig. 4). These results correlate with the observations from SEM and TEM measurements. Calculation of the form factor shows that the particle size and shape do not change up to 400$^{\circ}$C.

\begin{figure*}[htbp]
  \centering
  \includegraphics[width=0.98\textwidth]{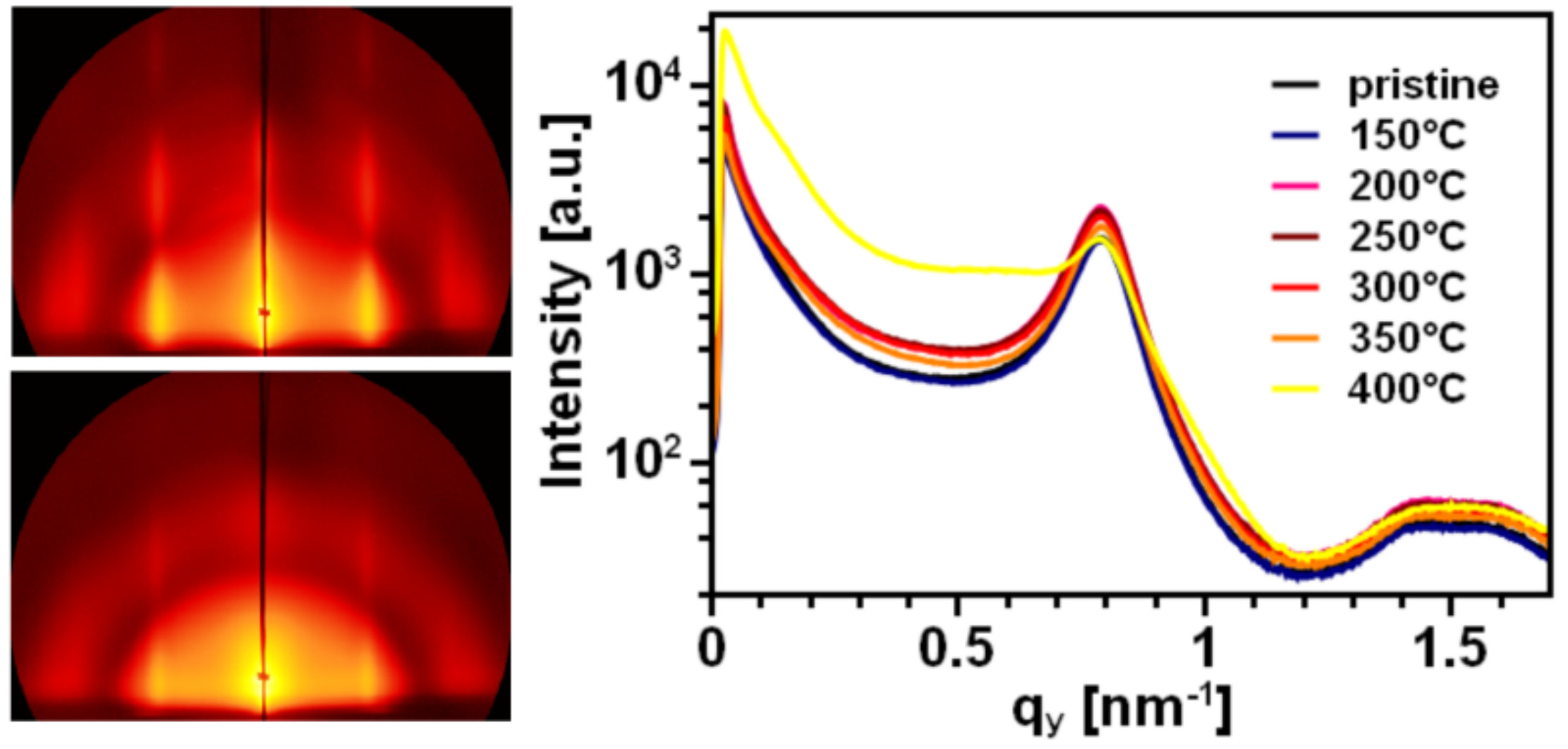}
  \caption{\textit{Left: GISAXS pattern of thermally untreated particle film (upper image) and a film annealed at 400$^{\circ}$C (lower image). The structural change is well visible in the loss of clarity of the reflections. Right: curves obtained by slicing the GISAXS pattern along the q$_{y}$ axis.}}
\end{figure*}

We investigated the behavior of the ligands during the annealing procedure by thermogravimetric analysis (TGA). Obviously the mass loss occurs stepwise as depicted in Fig. 5A. The first derivative of the measurement shows a first, small loss at 100$^{\circ}$C, attributed to water evaporation, a second, sharp signal at 220$^{\circ}$C, and a broader signal between 300 and 400$^{\circ}$C, until finally most of the ligands are lost above 400$^{\circ}$C. These signals are ascribed to the detachment of hexadecylamine and 1-adamantan carboxylic acid of the particle surface according to their binding energy \cite{28}. The loss of almost all ligand mass above 400$^{\circ}$C correlates with the coalescence of the particles which was observed by TEM and SEM.

\begin{figure}[htbp]
  \centering
  \includegraphics[width=0.45\textwidth]{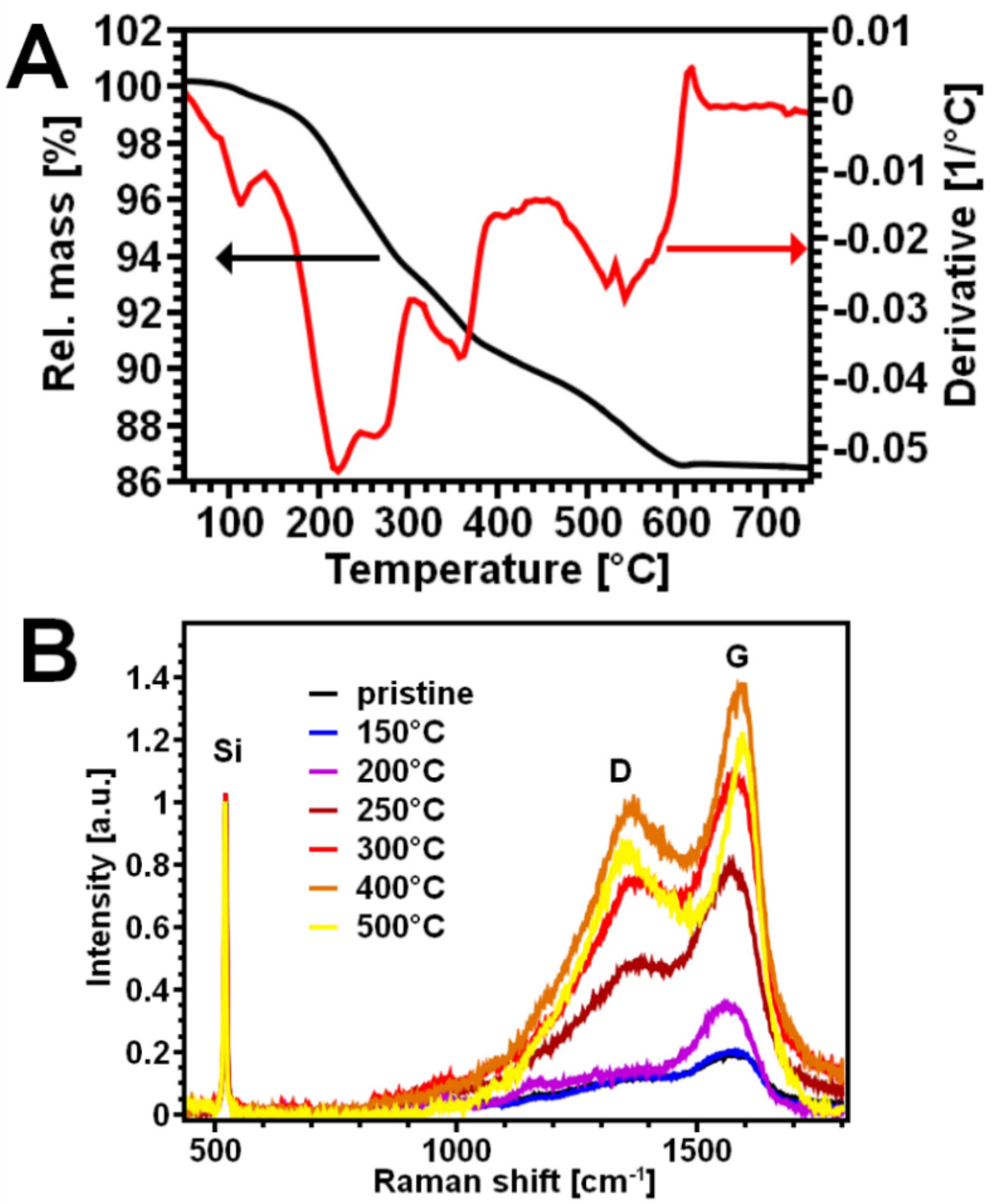}
  \caption{\textit{A) Result of a thermo-gravimetric measurement (black) and the first derivative (red). The measurement was performed under nitrogen flow and with a heating rate of 20$^{\circ}$C/min. B) Raman spectrum of a film annealed at different temperatures. The graphs are normalized to the silicon peak. The D and G peak intensity increases with rising annealing temperature. The G peak shifts to higher values with increasing annealing temperature.}}
\end{figure}

The measurements so far could explain the collapse of conductivity above 400$^{\circ}$C, but not the strong increase well below that temperature. In order to clarify this point we performed Raman spectroscopy on thermally treated and untreated nanoparticle films. The ligand shells around the nanoparticles consist of hydrocarbon molecules. For the analysis of carbon structures Raman spectroscopy is a common and versatile technique. Using visible laser light, a Raman spectrum between 1200 and 1800 cm$^{-1}$ is dominated by the sp$^{2}$-bonded carbon due to the preferred excitation of pi states. The spectrum depends strongly on the ordering of the sp$^{2}$ sites. Ferrari and Robertson \cite{29} introduced a model (FR model) to correlate the spectral features, i.e. D and G peaks, to the carbon nanostructure. The G peak around 1560 cm$^{-1}$ originates from the in-plane bond-stretching motion of sp$^{2}$-bonded carbon atoms. This relative motion can occur for all sp$^{2}$-bonded carbon atoms in rings and chains. The D peak around 1360 cm$^{-1}$ is forbidden for perfect graphite and occurs in the presence of disorder. For amorphous carbon both peaks are broad and overlap. Fig. 5B shows the Raman spectrum of the monolayer film after different annealing temperatures. The spectrum exhibit two broad overlapping features, namely, the D peak around 1360 cm$^{-1}$ and the G peak at around 1560-1590 cm$^{-1}$. Annealing up to 400$^{\circ}$C results in an increase of the D and G peak intensity and a shift of the G peak position to higher relative wave numbers \cite{30}. This is in accordance with increasing order of the carbon phase and is a result of stronger clustering of sp$^{2}$ carbon in aromatic rings and a shortening of the average C=C distance \cite{29}. The intensity increase of the G peak originates from the formation of conductive sp$^{2}$ carbon atoms resulting in the higher conductivity. At annealing temperature above 250$^{\circ}$C the D peak also starts to rise and indicates the generation of non conductive sp$^{3}$ carbon atoms, which could decrease the conductivity. Above 400$^{\circ}$C the D and G peak start to decrease, probably caused by the diffusion of carbon into the nanoparticles and the formation of cobalt carbides. With the breakdown of the sp$^{2}$ chains the conductivity starts to drop. At 500$^{\circ}$C annealing temperature a rest conductivity remains which might be due to a continuous film of carbon. The low conduction takes place in the carbon film and is not mediated by transport from nanoparticle to nanoparticle.

Below a temperature of 400$^{\circ}$C we do not observe any structural changes but a carbonization of the ligands remaining in between the nanoparticles. As a result the conductivity is elevated. At annealing temperatures above 400$^{\circ}$C the film collapses due to the loss of stabilizing ligands around the nanoparticles followed by coalescence to larger crystals. The interparticle distance becomes too large to conduct an electrical current.

\begin{figure}[htbp]
  \centering
  \includegraphics[width=0.45\textwidth]{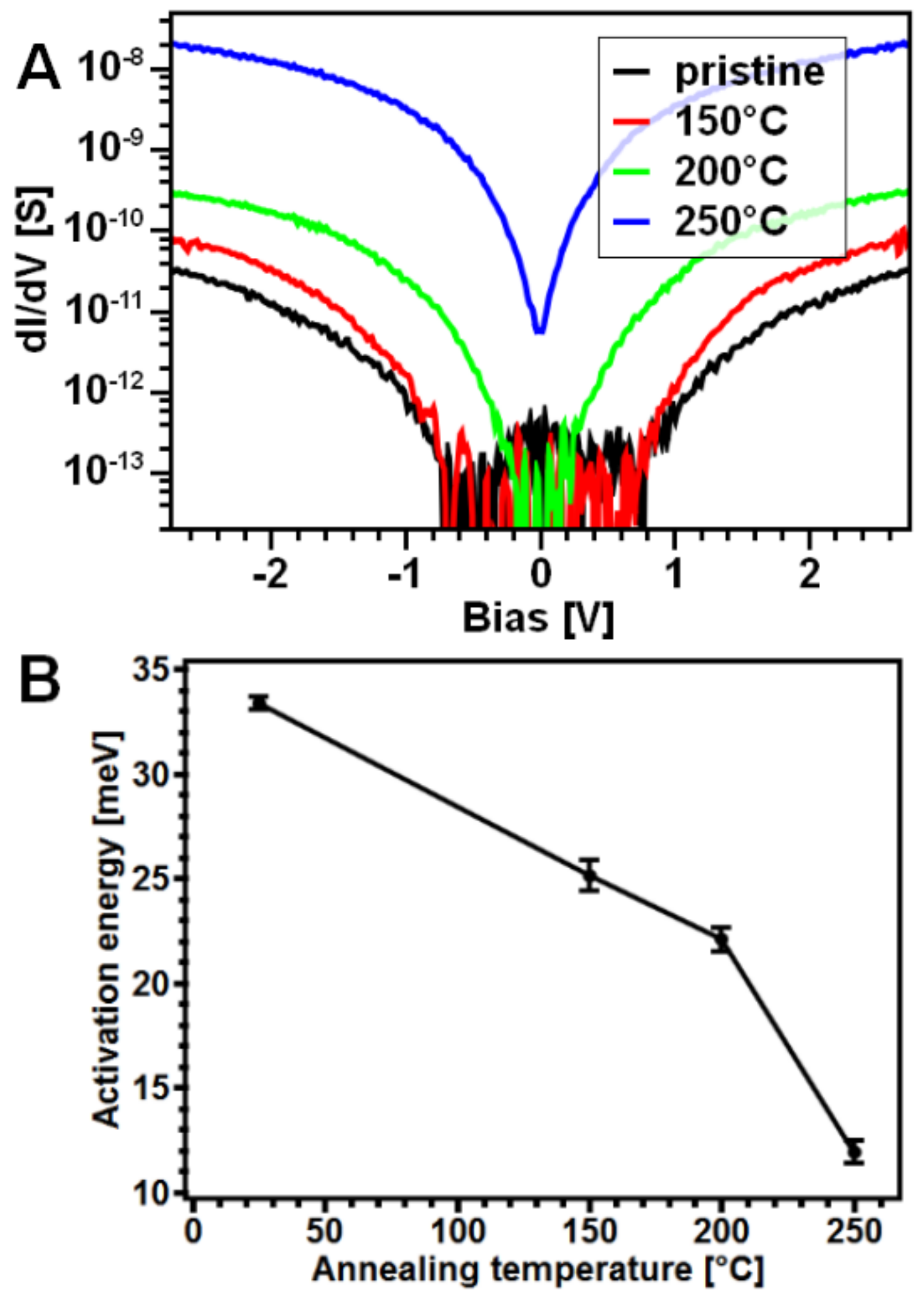}
  \caption{\textit{A) Differential conductance of a pristine and annealed films at 4.3 K. The film annealed at 250$^{\circ}$C shows no sign of Coulomb blockade at 4.3 K and zero bias, while the untreated film clearly shows blockade below half a volt. The film treated at 200$^{\circ}$C shows reduced Coulomb blockade. B) The activation energy decreases with increasing annealing temperature (at 2 V).}}
\end{figure}

In a previous paper \cite{11} we presented investigations on the transport properties of CoPt nanoparticle monolayer films at low temperature. At low temperatures nanoparticles show current suppression known as Coulomb blockade \cite{31} which originates from a charging energy $E_{C}$ of the particles due to their small capacitance. The total capacitance of the particles $C_{\Sigma}$ is the sum of the self-capacitance $C_{S}$ and the contact capacitance of a particle to the adjacent particles $C_{geo}$: $E_{C} = e^{2}/2C_{\Sigma}$ with $C_{\Sigma} = C_{S}+C_{geo}$. The observation of the Coulomb blockade requires the thermal energy to be lower than the particle's charging energy: $E_{C} \gg k_{B}T$. Monolayer films without thermal treatment show Coulomb blockade at temperatures below 40 K. For T < 40 K the electrons are tunneling when the applied field strength is sufficient to overcome the particle's charging energy. Above 40 K the electrons are thermally excited and thermally activated hopping can be expected \cite{11}. For monolayers which suffered a thermal treatment (above 250$^{\circ}$C) we do not observe Coulomb blockade. Figure 6A compares the conductance (at 4.3 K) of monolayer films without thermal treatment and of films annealed at 150$^{\circ}$C, 200$^{\circ}$C, and 250$^{\circ}$C. While the untreated film clearly shows a Coulomb blockade characteristic, the film annealed at 250$^{\circ}$C has a conductivity of about 10-11 S at 4.3 K and zero bias. The sample thermally treated at 200$^{\circ}$C shows a decreased Coulomb blockade voltage and an increased conductance compared to the untreated film. The high zero bias conductivity of the film annealed at 250$^{\circ}$C might originate from the transformation of the insulating ligand shell to conductive sp$^{2}$-hybridized carbon atoms. As a result the tunnel barrier of the particles and thus the tunnel resistant decreases. The electrons can overcome the tunnel barrier easier and lower charging energy is required due to stronger coupling of the particle to the adjacent nanocrystals. Figure 6B shows the activation energy plotted against the annealing temperature of the film. The activation energies are deducted from the slope of a corresponding Arrhenius plot. They decrease with increasing annealing temperature. For the particles without thermal treatment an activation energy of 33.56 meV is obtained at 2 V. This result is consistent with the results we presented in a previous publication \cite{11}. The particles annealed at 150$^{\circ}$C showed an activation energy of 25.64 meV. For the particles treated at 200$^{\circ}$C the activation energy is 22.50 meV and at 250$^{\circ}$C a value of 11.55 meV is obtained. These results demonstrate that the activation energy depends not only on the particle size and the applied field, but the coupling to the environment plays a major role.

\subsection*{CONCLUSION}

To conclude, by simple thermal annealing we managed to tune the electrical conductivity, the Coulomb blockade, and the activation energy of monolayers of monodisperse nanoparticle films. The monolayer films as prepared clearly exhibit Coulomb blockade, while annealing at 250°C is sufficient to suppress this phenomena. The investigated nanoparticle films showed a decrease in activation energy with increasing annealing temperature. The results show that the activation energy is not only influenced by the particle size and the applied field but also by the coupling to the environment. A first attempt of explanation, a nanoparticle displacement, could not be confirmed for annealing temperatures below 400°C. A second approach, a carbonization of the ligands, however, could be confirmed by Raman spectroscopy. The tunability of the coupling between the particles is of major importance in the development of advanced nanoparticle based devices.

\subsection*{ACKNOWLEDGEMENTS}

Financial support from the Free and Hanseatic City of Hamburg in the context of the "Landesexzellenzinitiative Hamburg: Spintronics" is gratefully acknowledged.  

\subsection*{EXPERIMENTAL DETAILS}

\textit{Chemicals}

Toluene and 2-propanol (both p.a. Merck), diphenyl ether (DPE, 99\%, Alfa Aesar), 1,2-hexadecandiol (HDD, 90\%, Fluka), 1-adamantanecarboxylic acid (ACA, 99\%, Fluka), 1,2-dichlorobenzene (99\%, Acros Organics), cobalt carbonyl (Co$_{2}$(CO)$_{8}$, stabilised with 1-5\% of hexane, Strem), and platinum(II)-acetylacetonate (Pt(acac)$_{2}$, 98\%, Strem), hexadecylamine (HDA, Merck) were of the highest purity available and used as received.

\

\textit{Nanoparticle synthesis}

The synthesis was carried out by using standard Schlenk line technique under nitrogen. CoPt$_{3}$ nanoparticles were synthesized via simultaneous reduction of platinum acetylacetonate (Pt(acac)$_{2}$) and thermal decomposition of cobalt carbonyl (Co$_{2}$(CO)$_{8}$) in the presence of 1-adamantan carboxylic acid (ACA) and hexadecylamine (HDA) as stabilizing agents, following a method similar to the one introduced by Shevchenko et al. \cite{26}. In a standard procedure HDA (8.0 g), ACA (0.496 g), HDD (0.26 g) and Pt(acac)$_{2}$ (65.6 mg) were dissolved in diphenylether (4 mL) at 65$^{\circ}$C. Once a optical clear solution was obtained, the mixture was degassed 3 times and heated to 155$^{\circ}$C and the Co$_{2}$(CO)$_{8}$ (92 mg) dissolved in 1,2-dichlorobenzene (1.6 mL) was injected quickly, under rigorous stirring. The mixture was stirred at the injection temperature for one hour and then the temperature was increase to 230$^{\circ}$C for two hours. Obtained particles were precipitated with 2-propanol, centrifuged and redissolved in toluene. This procedure was repeated once in order to remove excess ligands from the solution. At the end, the nanoparticle solution was filtered through a PTFE 0.2 $\mu$m filter.

\

\textit{Langmuir-Blodgett for monolayer films}

The monolayer films were prepared in a Langmuir-Blodgett trough 311D from NIMA Technology following the method developed by Aleksandrovic et al. \cite{10}. The nanoparticles were washed one more time and redissolved in toluene. Diethylenglycol was used as subphase and the nanoparticles were compressed with 6 mm/min to a target pressure of 10 mN/m. The monolayer film was hold at the target pressure for about two hours to allow the film to relax and reorganized. Afterwards the monolayer film was deposit on a Si/SiO$^{2}$-wafer surface which was structured with gold electrodes by e-beam lithography. The annealing of the monolayer films were carried out under vacuum (30 min) to avoid the oxidation of cobalt. 

\

\textit{Electrical measurements}

The room temperature electrical measurements were performed with an Agilent 4156C precision semiconductor parameter analyzer. For the low temperature electrical measurements we used a 4200-SCS semiconductor characterization system from Keithley Instruments and the VFTTP4 probestation by Lake Shore Cryotronics. 

\

\textit{GISAXS}

The GISAXS measurements were performed with the grazing incidence setup of the experimental station A2 at HASYLAB (Hamburg, Germany) equipped with a high-resolution 2-D CCD detector (MAR research, 2048 x 2048 pixel, pixel size 79 $\mu$m). The bending magnet beamline has a fixed wavelength of 0.15 nm and a distance of 2 m between sample and detector. The sample was placed horizontally on the goniometer and tilted to a glancing incidence angle of 0.6$^{\circ}$, which is well above the critical angle of the native oxide layer of the silicon substrate. The accumulation time was 300 s. Out-of-plane cuts (Intensity vs. q$_{y}$ at constant q$_{z}$; dq$_{z}$ = 0.01 nm$^{-1}$) were made at the Yoneda maximum alpha$_{f}$ = alpha$_{c}$ = 0.23$^{\circ}$).

\

\textit{Raman spectroscopy}

The experiments were performed using the Ar-Kr laser line 514.5 nm (2.41 eV) with a high numerical aperture Zeiss Epiplan Apochromat objective (150x, NA = 0.9). The laser power was 2.5 mW corresponding to 2 MW/cm$^{2}$ while the grating employed had 1200 groves/mm.

\end{document}